\documentclass[aps,prd,nofootinbib,twocolumn,superscriptaddress,floatfix,notitlepage]{revtex4-1}

\pdfoutput=1
\usepackage{graphicx}
\usepackage{amsmath,amsfonts,amssymb}
\usepackage{color}
\usepackage[breaklinks,colorlinks,urlcolor=blue,citecolor=blue,linkcolor=magenta]{hyperref}
\usepackage{verbatim}
\usepackage{enumitem}
\usepackage{braket}
\usepackage{hyperref}
\usepackage{aas_macros}

\newcommand{\be}{\begin{equation}} 
\newcommand{\ee}{\end{equation}}
\newcommand{\bea}{\begin{equation}\begin{aligned}} 
\newcommand{\eea}{\end{aligned}\end{equation}}
\newcommand{\td}{{\rm d}}
\newcommand{\Msun}{M_\odot}

\def\lsim{\mathrel{\raise.3ex\hbox{$<$\kern-.75em\lower1ex\hbox{$\sim$}}}}
\def\gsim{\mathrel{\raise.3ex\hbox{$>$\kern-.75em\lower1ex\hbox{$\sim$}}}}

\begin{document}

\title{Prospects for Future Binary Black Hole GW Studies in Light of PTA Measurements}

\author{John Ellis}
\email{john.ellis@cern.ch}
\affiliation{King’s College London, Strand, London, WC2R 2LS, United Kingdom}
\affiliation{Theoretical Physics Department, CERN, Geneva, Switzerland}
\affiliation{Keemilise ja Bioloogilise F\"u\"usika Instituut, R\"avala pst. 10, 10143 Tallinn, Estonia}
\author{Malcolm Fairbairn}
\email{malcolm.fairbairn@kcl.ac.uk}
\affiliation{King’s College London, Strand, London, WC2R 2LS, United Kingdom}
\author{Gert H\"utsi}
\email{gert.hutsi@kbfi.ee}
\affiliation{Keemilise ja Bioloogilise F\"u\"usika Instituut, R\"avala pst. 10, 10143 Tallinn, Estonia}
\author{Martti Raidal}
\email{martti.raidal@cern.ch}
\affiliation{Keemilise ja Bioloogilise F\"u\"usika Instituut, R\"avala pst. 10, 10143 Tallinn, Estonia}
\author{Juan Urrutia}
\email{juan.urrutia@kbfi.ee}
\affiliation{Keemilise ja Bioloogilise F\"u\"usika Instituut, R\"avala pst. 10, 10143 Tallinn, Estonia}
\affiliation{Departament of Cybernetics, Tallinn University of Technology, Akadeemia tee 21, 12618 Tallinn, Estonia}
\author{Ville Vaskonen}
\email{ville.vaskonen@pd.infn.it}
\affiliation{Keemilise ja Bioloogilise F\"u\"usika Instituut, R\"avala pst. 10, 10143 Tallinn, Estonia}
\affiliation{Dipartimento di Fisica e Astronomia, Universit\`a degli Studi di Padova, Via Marzolo 8, 35131 Padova, Italy}
\affiliation{Istituto Nazionale di Fisica Nucleare, Sezione di Padova, Via Marzolo 8, 35131 Padova, Italy}
\author{Hardi Veerm\"ae}
\email{hardi.veermae@cern.ch}
\affiliation{Keemilise ja Bioloogilise F\"u\"usika Instituut, R\"avala pst. 10, 10143 Tallinn, Estonia}

\begin{abstract}
NANOGrav and other Pulsar Timing Arrays (PTAs) have discovered a common-spectrum process in the nHz range that may be due to gravitational waves (GWs): if so, they  are likely to have been generated by black hole (BH) binaries with total masses $> 10^9 \Msun$. 
Using the Extended Press-Schechter formalism to model the galactic halo mass function and a simple relation between the halo and BH masses suggests that these binaries have redshifts $z = {\cal O}(1)$ and mass ratios $\gtrsim 10$,
and that the GW signal at frequencies above ${\cal O}(10)$~nHz may be dominated by relatively few binaries that could be distinguished experimentally and would yield observable circular polarization. 
Extrapolating the model to higher frequencies indicates that future GW detectors such as LISA and AEDGE could extend the PTA observations to lower BH masses $\gtrsim 10^3 \Msun$.

\vspace{3mm}
\noindent
KCL-PH-TH/2023-04, CERN-TH-2023-008, AION-REPORT/2023-1
\end{abstract}

\maketitle

%%%%%%%%%%%%%%%%%%%%%%%%%%%%%%%%%%%%%%%%%%%
\section{Introduction}
%%%%%%%%%%%%%%%%%%%%%%%%%%%%%%%%%%%%%%%%%%%
The discovery~\cite{LIGOScientific:2016aoc} by the LIGO and Virgo experiments of gravitational waves (GWs) in the range of frequencies ${\cal O}(100)$~Hz generated by mergers of black holes (BHs) with masses ${\cal O}(10 - 100) \Msun$ ~\cite{LIGOScientific:2021psn}, has opened a new window on both astrophysics and cosmology. 
Supermassive black holes (SMBHs) with masses ${\cal O}(10^6 - 10^{10})\Msun$ are known to be present in galactic nuclei, and the immediate surroundings of two of them have recently been imaged by the Event Horizon Telescope~\cite{EventHorizonTelescope:2019dse,EventHorizonTelescope:2022wkp}, but information about intermediate-mass black holes (IMBHs) with masses in the range ${\cal O}(10^3 - 10^5) \Msun$~\cite{Greene:2019vlv} is less complete.
Observations of IMBHs and their mergers would cast light on the uncharted mechanisms which must exist in order for SMBHs to form.

A prerequisite for BH mergers is the formation of tightly-bound binaries that can radiate GWs efficiently. While there are several well-understood paths for forming compact stellar-mass BHs binaries (e.g., common envelope evolution or dynamical capture in dense stellar environments) the final stage of SMBH binary formation in galaxy mergers is still not fully understood and is commonly known as ``the final parsec problem''~\cite{Begelman:1980vb}. 
However, observations of quasar optical variability provide indirect evidence for the existence of some tight binary SMBHs in the regime where the emission of GWs must already have a noticeable effects on their orbital evolution~\cite{Valtonen:2008tx,Boroson:2009va,Iguchi:2010zs,Graham:2015gma,ONeill:2021swa}. 
GW data will be essential for a better understanding how SMBHs overcome the the ``final parsec'' obstacle, and how the assembly of the SMBH population proceeds in general.

Pulsar timing arrays (PTAs) are potentially sensitive to GWs in the nHz range, and
NANOGrav and other PTAs have recently reported evidence for a common-spectrum stochastic process~\cite{NANOGrav:2020bcs,Goncharov:2021oub,Chen:2021rqp,Antoniadis:2022pcn}.
Their signals have power spectra that are consistent with predictions based on inspiralling binary SMBH models~\cite{Phinney:2001di},~\footnote{A similar signal could be generated by primordial BHs~\cite{Vaskonen:2020lbd, DeLuca:2020agl, Kohri:2020qqd, Ashoorioon:2022raz}, but might require modifications of models based on simple cosmic string networks~\cite{Ellis:2020ena,Blasi:2020mfx,Buchmuller:2020lbh}: see, e.g., \cite{Blanco-Pillado:2021ygr}.} but they have not (yet) detected the  Hellings-Downs quadrupolar signature that is characteristic of GWs~\cite{Hellings:1983fr}.

The BH merger interpretation of the PTA measurements does
not provide direct information on the masses of the infalling
BHs, which requires modelling and measurements of this
stochastic GW background in different frequency ranges~\cite{Sesana:2008mz}.
The purpose of this paper is to illustrate how 
future measurements of the GW spectrum due to unresolved infall sources and individual binary mergers
at frequencies between the PTAs and LIGO/Virgo
will be able to extend the PTA measurements
to lower black-hole mass ranges, probing
models for the assembly of SMBHs, and
perhaps obtaining direct evidence for the mergers of IMBHs.
We take as the amplitude of the common-spectrum process measured by the PTAs
$A = 2.8^{+1.2}_{-0.8} \times 10^{-15}$ at a reference frequency of 1/yr, as
found by the International Pulsar Timing Array (IPTA)~\cite{Antoniadis:2022pcn} assuming the spectral index of $\alpha = -2/3$
expected from inspiralling SMBHs, which is
consistent with other measurements.

We use as examples of next-generation GW detectors the planned LISA
laser interferometer~\cite{Berti:2005ys}, whose peak sensitivity is at
frequencies ${\cal O}(10^{-4} - 10^{-2})$~Hz, the proposed ET laser interferometer~\cite{Sathyaprakash:2012jk}, whose peak sensitivity is at
frequencies ${\cal O}(1 - 10^{2})$~Hz and the
projected AION-km and AEDGE atom interferometers~\cite{Badurina:2019hst,AEDGE:2019nxb}, whose peak
sensitivities would be in the range ${\cal O}(10^{-2} - 1)$~Hz.
We show that, whereas none of these detectors would
observe a signal from the binary black holes weighing $> {\cal O}(10^9)\Msun$ that would probably be responsible for most of the PTA signal, 
LISA would observe a merger signal from
binary black holes weighing ${\cal O}(10^{3} - 10^{9})\Msun$ that might contribute part of the PTA signal,
while AEDGE would observe a merger signal if there is a population of 
binary black holes weighing ${\cal O}(10^{3} - 10^{6})\Msun$ with a formation history similar to the heavier BHs responsible for the PTA signal.
%, and AION-km might also be able to observe some mergers.

%\textcolor{red}{$\Omega_{\rm yr} = 2\pi^2 A_{\rm yr}^2 f_{\rm yr}^2/(3H_0^2)$ where $A_{\rm yr} = 2.8^{+1.2}_{-0.8} \times 10^{-15}$~\cite{Antoniadis:2022pcn}}

Throughout this paper we use natural units with $c=1$, $G_N = 1$.

%%%%%%%%%%%%%%%%%%%%%%%%%%%%%%%%%%%%%%%%%%%
\section{BH merger rate}
%%%%%%%%%%%%%%%%%%%%%%%%%%%%%%%%%%%%%%%%%%%

As a first step towards estimating the BH merger rate,
we use the Extended Press-Schechter (EPS) formalism~\cite{Press:1973iz,Bond:1990iw} to calculate the galactic halo mass function,
\be
	\frac{\td n(M,t)}{\td \ln M} = \frac{\rho_0}{M} \sqrt{\frac{2}{\pi}} \frac{\td \ln \sigma}{\td \ln M} \frac{\delta_c(z)}{\sigma (M)}e^{-\frac{\delta_c(z)^2}{2\sigma^2(M)}} \, ,
\ee
where $\rho_0$ is the background matter density today, $\sigma^2(M)$ is the variance of the matter fluctuations, $\delta_c(z) \approx 1.686/D(z)$ where $D(z)$ is the linear growth function~\cite{Dodelson:2003ft},
and the probability per unit time for a halo of mass $M_1$ to merge with another one of mass $M_2$ at some redshift $z$ and become a halo of mass $M_f = M_1+ M_2$ is~\cite{Lacey:1993iv}
\bea
	&\frac{\td p(M_1,M_2,t)}{\td t \td M_2} = \frac{1}{M_f}\sqrt{\frac{2}{\pi}}\left|\frac{\dot{\delta}_c}{\delta_c}\right| \frac{\td\ln{\sigma}}{\td\ln{M_f}} \left[1-\frac{\sigma^2(M_f)}{\sigma^2(M_1)}\right]^{-\frac{3}{2}} \\
	&\times  \frac{\delta_c}{\sigma(M_f)}  \exp\!\left[-\frac{\delta_c^2}{2}\left(\frac{1}{\sigma^2(M_f)}-\frac{1}{\sigma^2(M_1)}\right)\right] \,.
\eea
We calculate $\sigma^2(M)$ from the CDM matter power spectrum with the Planck 2018 cosmological parameters~\cite{Planck:2018vyg} and the transfer function derived in~\cite{Eisenstein:1997ik}. Defining the halo merger rate kernel
\be
	Q(M_1,M_2,t) \equiv \frac{\td p(M_1,M_2,t)}{\td t \td M_2}  \left[\frac{\td n(M_2,t)}{\td M_2} \right]^{-1} ,
\ee
the halo merger rate is
\be
	\frac{\td R_h}{\td M_1 \td M_2} = \frac{\td n(M_1,t)}{\td M_1} \frac{\td n(M_2,t)}{\td M_2} Q(M_1,M_2,t) \,.
\ee
We note that $Q(M_1,M_2,t)$ is not symmetric in the exchange of $M_1$ and $M_2$. 
It was shown in~\cite{Erickcek:2006xc} that taking $M_1<M_2$ in the merger kernel approximates well the result of Benson-Kamionkowski-Hassani (BKH) merger theory~\cite{Benson:2004gv} where the merger rate kernel preserves the EPS halo mass function. We will, therefore, require that the first argument of $Q$ is smaller than the second.

In order to estimate the merger rate of central BHs, we must estimate the probability $p_{\rm occ}(m|M)$ that a BH of mass $m$ occupies a halo of mass $M$, as well as the probability $p_{\rm merg}(m_1,m_2)$ that the galactic merger leads to a merger of their central BHs. The resulting BH merger rate can be expressed as
\bea \label{eq:BHmergerrate}
    \frac{\td R_{\rm BH}}{\td m_1 \td m_2} =
    &\int \td M_1 \td M_2 \, p_{\rm merg}(m_1,m_2) \\ 
    & \times p_{\rm occ}(m_1|M_1) p_{\rm occ}(m_2|M_2) \frac{\td R_h}{\td M_1 \td M_2} \,.
\eea
We assume the following simple redshift-dependent relation between the halo mass and the BH mass~\cite{Barkana:2000fd, Wyithe:2002ep}:
\be\label{mbhmh}
    \frac{M_v}{10^{12}M_\odot} 
    \!=\! 10.5 \left[ \frac{\Omega_M(0)}{\Omega_M(z)} \frac{\Delta_c(z)}{18\pi^2} \right]^{\!-\frac12} \!\!(1+z)^{-\frac32} \!\left[\frac{m_{\rm BH}}{10^8 M_\odot}\right]^{\!\frac35} \,,
\ee
where $\Delta_c(z) = 18 \pi^2 + 82 [\Omega_M(z)-1] - 39 [\Omega_M(z)-1]^2$ is the critical overdensity at virialization, which corresponds to $p_{\rm occ}(m|M) = p_{\rm occ}(m) \delta(m - m_{\rm BH}(M))$. It is generally expected that SMBHs inhabit most large galaxies, and X-ray observations~\cite{Miller:2014vta} constrain
the SMBH occupation fraction $p_{\rm occ}(m)$ to be $>20\%$ for early galaxies with lower stellar masses $10^{7} < M_{*}/\Msun < 10^{10}$. 

\begin{figure}
\centering
\includegraphics[width=0.96\columnwidth]{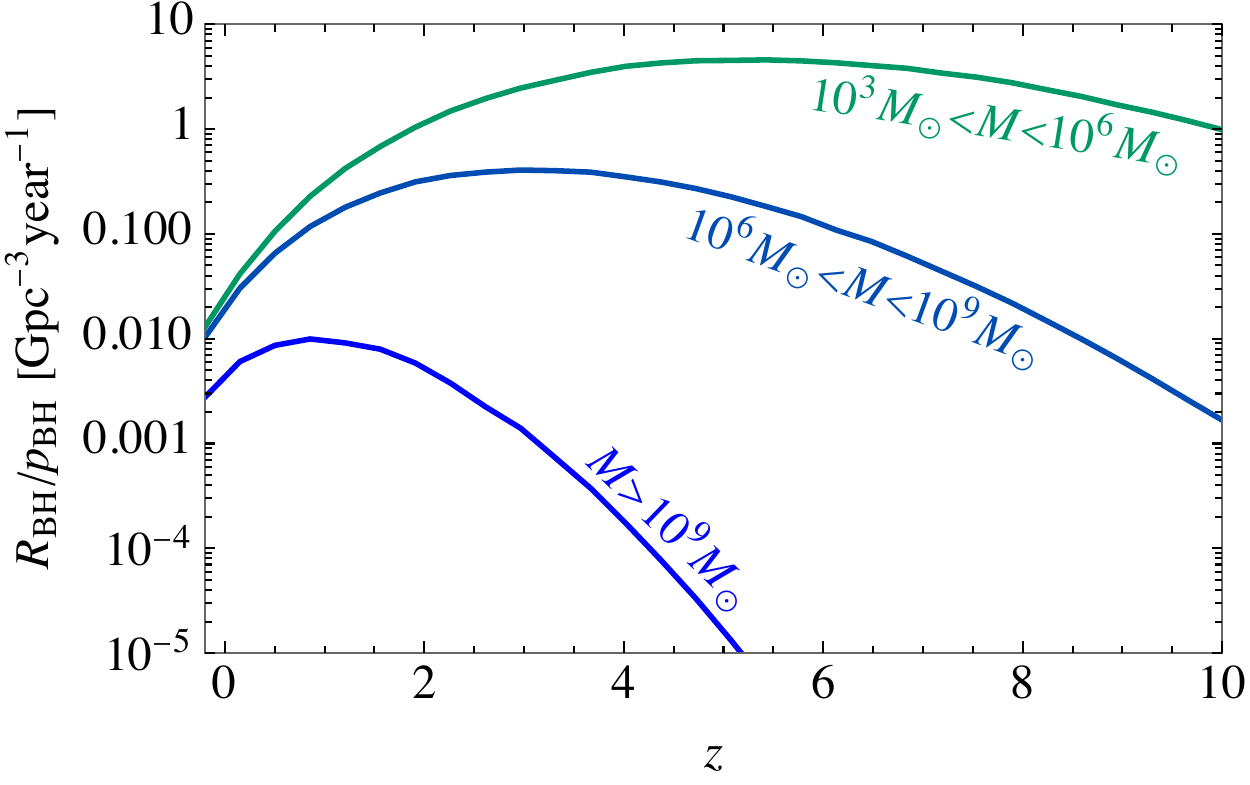}
\caption{The BH merger rates calculated as functions of the 
redshift $z$ in three ranges of the
total merging BH mass $M$ and normalized relative to the merger
probability $p_{\rm BH}$ discussed in the text.}
\label{fig:RBH}
\end{figure}

\begin{figure*}
\centering
\includegraphics[width=1.3\columnwidth]{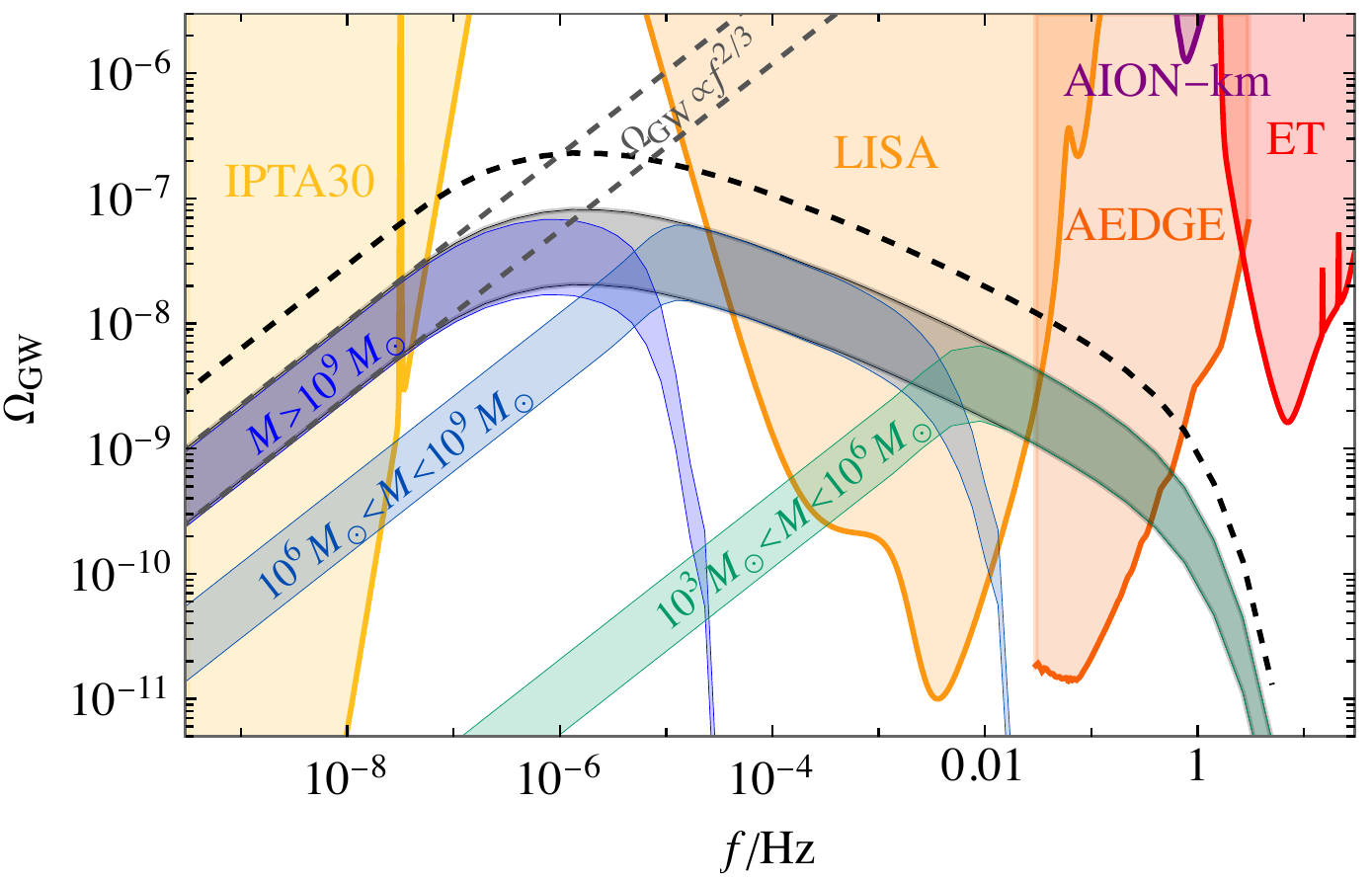}
\caption{The mean GW energy density spectrum from massive BH mergers compared with the sensitivities of different experiments. The black dashed curve shows the case where all mergers of galaxies produce a GW signal for BHs heavier than $10^3 \Msun$ ($p_{\rm BH} = 1$). The colored bands show the spectra from SMBHs heavier than $10^9 \Msun$ (dark blue), from SMBHs in the range $(10^6 \Msun,10^9 \Msun)$ (light blue), and from IMBHs with masses in the range $10^3-10^6 \Msun$ (green), assuming a universal efficiency factor $p_{\rm BH} = 0.17^{+0.18}_{-0.08}$. The shaded regions show the prospective sensitivities of SKA~\cite{Janssen:2014dka}, LISA~\cite{LISA:2017pwj}, AEDGE~\cite{AEDGE:2019nxb,Badurina:2021rgt}, AION-km~\cite{Badurina:2019hst,Badurina:2021rgt} and ET~\cite{Sathyaprakash:2012jk}.}
\label{fig:SGWB}
\end{figure*}

In view of the paucity of information about the merger probability $p_{\rm merg}(m_1,m_2)$, we model $p_{\rm merg}$ by a constant for the sake of simplicity. After the galactic merger, various dynamical mechanisms must decrease the size of the SMBH binary below sub-parsec scales in order for a GW emission-driven merger to take place~\cite{Begelman:1980vb}. The crossing of the final parsec is determined mostly by binary hardening via stellar loss-cone scattering~\cite{Merritt:2013awa}, which can currently be considered the largest source of uncertainty in $p_{\rm merg}$. The numerical study~\cite{Kelley:2016gse}, based on a population of $10^6-10^{10} \Msun$ BH binaries in the {\tt Illustris} simulation~\cite{Vogelsberger:2014kha}, found for a wide range of model parameters that the coalescing fraction is nearly independent of the total mass of the binary but decreases with the mass ratio. Neglecting the latter dependence should not significantly affect our results, since extreme mass ratio inspirals contribute subdominantly to the GW background.
With these assumptions, we obtain
\be
    \frac{\td R_{\rm BH}}{\td m_1 \td m_2} 
    \approx p_{\rm BH} \frac{\td M_1}{\td m_1} \frac{\td M_2}{\td m_2} \frac{\td R_h}{\td M_1 \td M_2} \, ,
\ee
where $p_{\rm BH} \equiv p_{\rm occ}(m_1)p_{\rm occ}(m_2) p_{\rm merg}$ denotes the probability that the halo merger leads to a merger of the central SMBHs. 

We show in Fig.~\ref{fig:RBH} the BH merger rates as functions of the 
redshift $z$, calculated using this approach in three ranges of the
total merging BH mass $M$ and normalized relative to the merger
probability $p_{\rm BH}$ discussed above. We see that mergers with
total masses $M > 10^9 \Msun$ occur typically at $z = {\cal O}(1)$, 
those with
total masses $M \in (10^6, 10^9) \Msun$ occur typically at $z = {\cal O}(3)$,
and those with total masses $M \in (10^3, 10^6) \Msun$ occur typically at $z = {\cal O}(5)$.

In this paper, we use the IPTA measurement~\cite{Antoniadis:2022pcn} to normalize $p_{\rm BH}$ for large BH masses, which circumvents the astrophysical uncertainties related to $p_{\rm occ}$ and $p_{\rm merg}$, and extrapolate it to smaller BH masses. We also comment on the potential mass dependence of $p_{\rm BH}$, which is relevant for the GW phenomenology when we extrapolate from the frequency band relevant for PTAs to those to be explored by next-generation GW detectors. %addressed below by allowing $p_{\rm BH}$ to vary in different mass ranges. }

%%%%%%%%%%%%%%%%%%%%%%%%%%%%%%%%%%%%%%%%%%%
\section{Analysis}
%%%%%%%%%%%%%%%%%%%%%%%%%%%%%%%%%%%%%%%%%%%

\begin{figure*}
\centering
\includegraphics[width=1.8\columnwidth]{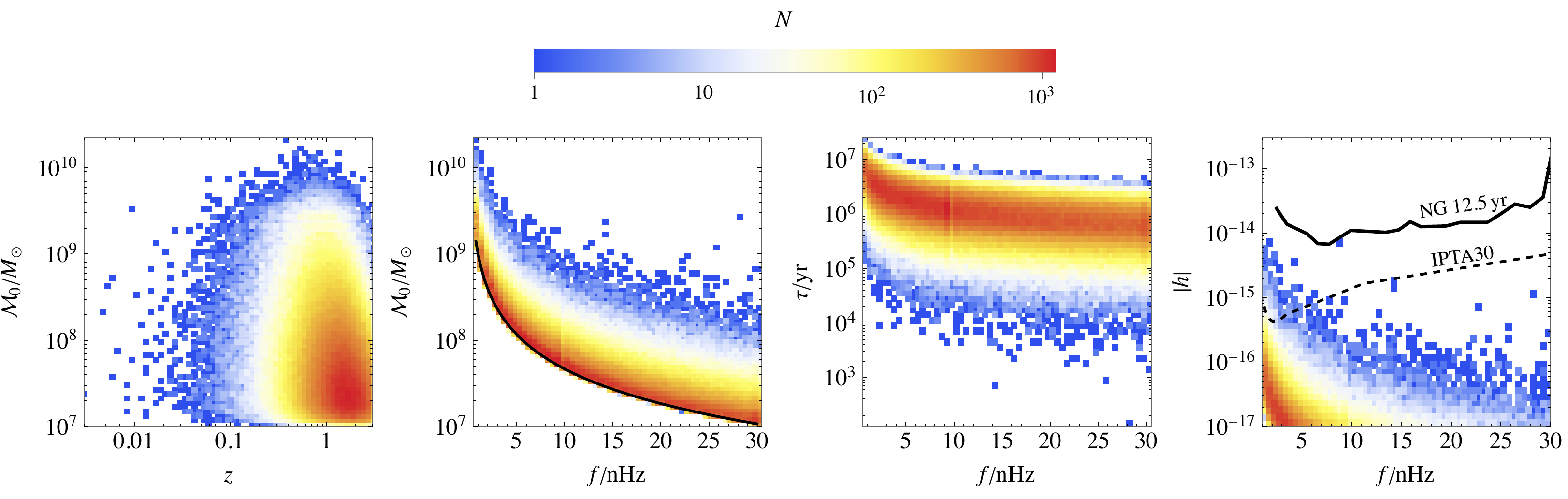}
\caption{The distributions of the binary parameters for a sampling of the BH binary population with $0<z<3$, $1\,{\rm nHz} < f < 30 {\rm nHz}$ and $\mathcal{M}_0 > 3\times 10^8 (f/{\rm nHz})^{-\frac43}$. The symmetric mass ratio is integrated with the lower bound $\eta > 0.001$.}
\label{fig:distr}
\end{figure*}

\subsection{The GW energy spectrum}

The mean GW energy density spectrum from the SMBH binary population can be estimated as~\cite{Phinney:2001di}
\be\label{eq:OmegaGW}
    \Omega_{\rm GW}(f) 
    \equiv \frac{1}{\rho_{\rm c}}\frac{\td \rho_{\rm GW}}{\td \ln f}
    = \int \text{d}\lambda \,\frac{2\pi}{5} \frac{f^3 |\tilde h(f)|^2}{\rho_c} \,,
\ee
where $|\tilde h(f)|$ denotes the optimal amplitude of the Fourier transform of the GW strain, $\rho_{\rm c} = 3H_0^2/8\pi$,
\be
    \td \lambda 
    = \td m_1 \td m_2 \,\frac{\td z}{1+z} \frac{\td V_c}{\td z} \frac{\td R_{\rm BH}(t)}{\td m_1 \td m_2} \, 
\ee
is the differential merger rate and $V_c$ denotes the comoving volume available at redshift $z$. We compute $|\tilde h(f)|$ using the inspiral-merger-ringdown template~\cite{Ajith:2007kx}
\bea\label{eq:A(f)}
    |\tilde h(f)| &= \sqrt{\frac{5}{24}}\frac{\mathcal{M}_z^{\frac{5}{6}}}{\pi^{\frac{2}{3}}D_L}\\
    &\times \begin{cases} 
      f^{-\frac{7}{6}} & f<f_{\rm merg} \\
      f_{\rm merg}^{-\frac12}f^{-\frac23} & f_{\rm merg}\leq f< f_{\rm ring} \\
      f_{\rm merg}^{-\frac12}f_{\rm ring}^{-\frac23}\frac{\sigma^2}{4(f-f_{\rm ring})^2+\sigma^2} & f_{\rm ring}\leq f<f_{\rm cut} \,,
    \end{cases}
\eea
where redshift dependent the chirp mass $\mathcal{M}_z$ is given in terms of the total mass $M$ and the symmetric mass ratio $\eta \equiv m_1 m_2/M^2$ of the binary by $\mathcal{M}_z \equiv (1+z) \mathcal{M}_0 = (1+z) M \eta^{\frac35}$, and $D_L$ denotes the luminosity distance of the binary. This implies the canonical $\Omega_{\rm GW} \propto f^{\frac23}$ scaling during the inspiral phase~\cite{Phinney:2001di}. The frequencies $f_{\rm merg}$, $f_{\rm ring}$, $f_{\rm cut}$ and $\sigma$ are parameterised as $f_j = \eta^{\frac53}(a_j \eta^2 + b_j \eta + c_j)/(\pi \mathcal{M}_z)$ where $a_j$, $b_j$ and $c_j$ are coefficients whose fitted values are given in Table~I of~\cite{Ajith:2007kx}. 

We comment briefly on potential uncertainties in the GW energy density spectrum. If orbital decay is driven by other processes in addition to GW emission, such as viscous drag, then the coalescence time is shortened and the total GW spectrum is suppressed by a factor of $T_{\rm tot}/T_{\rm GW}$, where $T_{\rm tot}$, $T_{\rm GW}$ denote the characteristic hardening timescales~\cite{Kocsis:2010xa, Kelley:2016gse}. Here we omit this effect, assuming that the signal is dominated by binaries for which the hardening is driven mainly by GW emission. Additionally, we consider only circular binaries, which, in the inspiral phase, radiate monochromatically at twice the orbital frequency. We note, however, that the GW spectrum of eccentric binaries would contain higher harmonics, with the fundamental harmonic (at the orbital frequency) becoming dominant at eccentricities $e > 0.4$~\cite{Taylor:2015kpa}. In addition, the GW luminosity would be enhanced by a factor of $(1 +73/24 e^2 + 37/96 e^4)/(1-e^2)^{\frac72}$~\cite{Peters:1963ux, Enoki:2006kj, Kelley:2017lek}.

Fig.~\ref{fig:SGWB} shows the GW energy density spectrum from BH binaries in different mass ranges. The black dashed curve corresponds to $p_{\rm BH} = 1$, i.e., the assumption that each galactic merger would produce a BH merger with $m_2 > m_1 > 10^3 \Msun$. This naive assumption leads to a spectrum exceeding that observed in the PTA band. 

We stress that the spectrum in Eq.~\eqref{eq:OmegaGW}, which is displayed in Fig.~\ref{fig:SGWB}, does not necessarily correspond to a stochastic GW background as it does not distinguish between individual resolvable events and unresolvable events that would contribute to a GW background. The $\Omega_{\rm GW} \propto f^{\frac23}$ tail at low frequencies arises from nearly monochromatic signals generated by inspiralling BH binaries. Because of this, $\Omega_{\rm GW}$ will fluctuate depending on the specific realisation of the binary population, so the spectrum in Fig.~\ref{fig:SGWB} should be interpreted as the mean $\Omega_{\rm GW}$ obtained by averaging over many potential realizations of the binary population. We return to the probability distribution of $\Omega_{\rm GW}$ in the next subsection and in Appendix~\ref{app:A}.

The grey band in Fig.~\ref{fig:SGWB} was obtained by choosing a fixed value of $p_{\rm BH}$ so that the amplitude of the total GW spectrum from BH binaries matches the IPTA observations in the nHz range, which requires $p_{\rm BH} = 0.17^{+0.18}_{-0.08}$.
The colored bands show the contributions to the total IPTA spectrum from different ranges of the BH masses. Comparing the dark blue, light blue and green bands we see that the largest contribution to the GW spectrum in the PTA window comes from $M > 10^9 \Msun$ SMBH binaries, consistent with earlier studies~\cite{Wyithe:2002ep,Sesana:2004sp,Enoki:2004ew,Sesana:2008mz,Sesana:2008xk,Kelley:2016gse,Izquierdo-Villalba:2021prf,Becsy:2022pnr}.

Some characteristics of these binaries are shown in Fig.~\ref{fig:distr}. The left panel shows that the total masses are $\lesssim 10^{10}$, and the middle panel shows that their redshifts are typically ${\cal O}(1)$.  The right panel shows that the binary lifetimes are only weakly correlated with their emission frequencies, with lifetimes $\gtrsim 10^5$~yr being generally favored.

Returning to Fig.~\ref{fig:SGWB}, we see that the GW energy spectrum in the LISA frequency range is dominated by lighter BH binaries in the mass range $M \in (10^6, 10^9) \Msun$, shown in light blue, and the contribution in the AEDGE band is dominated by IMBHs with masses in the range $M \in (10^3, 10^6) \Msun$, shown in green. The low-frequency GW emissions from the early stages of the inspirals of these binaries also contribute in the lower-frequency bands, but these contributions are subdominant compared to the signals from higher-mass SMBH mergers.~\footnote{In addition to the experiments shown in Fig.~\ref{fig:SGWB}, we have also considered the prospective sensitivities to GWs using the astrometric data from the Gaia and Nancy Grace Roman space telescopes~\cite{Wang:2020pmf}. Their nominal sensitivities lie well above the black dashed line in Fig.~\ref{fig:SGWB} corresponding to $p_{\rm BH} = 1$, but a possible improvement of the Roman sensitivity might enable the detection of SMBH binary signals at frequencies $\in (10^{-7}, 10^{-6})$~Hz. Moreover, while the prospective power-law integrated sensitivity to GWs through binary resonance by laser ranging of the Moon~\cite{Blas:2021mqw} reaches the gray band shown in Fig.~\ref{fig:SGWB}, the most likely signal at $f\sim 10^{-6}$\,Hz is well below that (see Sec.~\ref{sec:distribution}) and consists of a few nearly-monochromatic sources.}

We recall that the estimated spectrum in Fig.~\ref{fig:SGWB} assumes a constant $p_{\rm BH}$. Clearly, the relative contribution from binaries with lighter BH masses $< 10^9 \Msun$  could be  enhanced (reduced) by increasing (decreasing) $p_{\rm BH}$ for these binaries.
In order for mergers of BHs with masses in the range $(10^6, 10^9) \Msun$ to dominate the PTA signal $p_{\rm BH} \simeq 1$ would be required in this mass range, but even this maximal enhancement would be insufficient for mergers with $M < 10^6 \Msun$ to contribute significantly to the PTA signal. That said, we emphasise that the GW spectrum in either the LISA or AEDGE frequency range could reach the black dashed line if $p_{\rm BH} \simeq 1$ in the relevant mass range, and could also be enhanced if Population III stars make a significant contribution to the BH spectrum. Conversely, the spectrum could be suppressed in these ranges if $p_{\rm BH}$ is smaller than the value that fits the PTA data.

\subsection{Distribution of sources}
\label{sec:distribution}

The number of BH binaries at redshift $z$ emitting in a given frequency band can be estimated from the time the binary spends in that frequency band, which gives~\cite{Sesana:2008mz}
\be\label{eq:N}
    \frac{\td N}{\td \mathcal{M}_z \td \eta \td z \td \ln f} 
    = \frac{\td \lambda}{\td \mathcal{M}_z \td \eta \td z} \left| \frac{\td \tau}{\td \ln f} \right| \,,
\ee
where, assuming circular orbits and orbital decay via GW emission~\cite{Peters:1964zz}, the coalescence time of an inspiralling binary emitting GWs with frequency $(1+z)f$ is~\footnote{We remark that, in Eq.~\eqref{eq:N}, the observed time spent in a given frequency band is redshifted when compared to the time in the binary rest frame. This redshift has been accounted for in Eq.~\eqref{eq:tau(f)}.}
\be\label{eq:tau(f)}
    \tau(f) 
    = \frac{5}{256 \pi^{\frac83} \mathcal{M}_z^{\frac53}}  f^{-\frac83} \,.
\ee
%Eq.~\eqref{eq:N} works well for estimating the number of binaries as they emit nearly monochromatic GWs. However, if the time of emission within a given frequency band is not clearly defined, e.g., because the GW spectrum is not monochromatic, then a different approach must be used~\cite{Kelley:2017lek}.
\begin{figure}
\centering
\includegraphics[width=0.9\columnwidth]{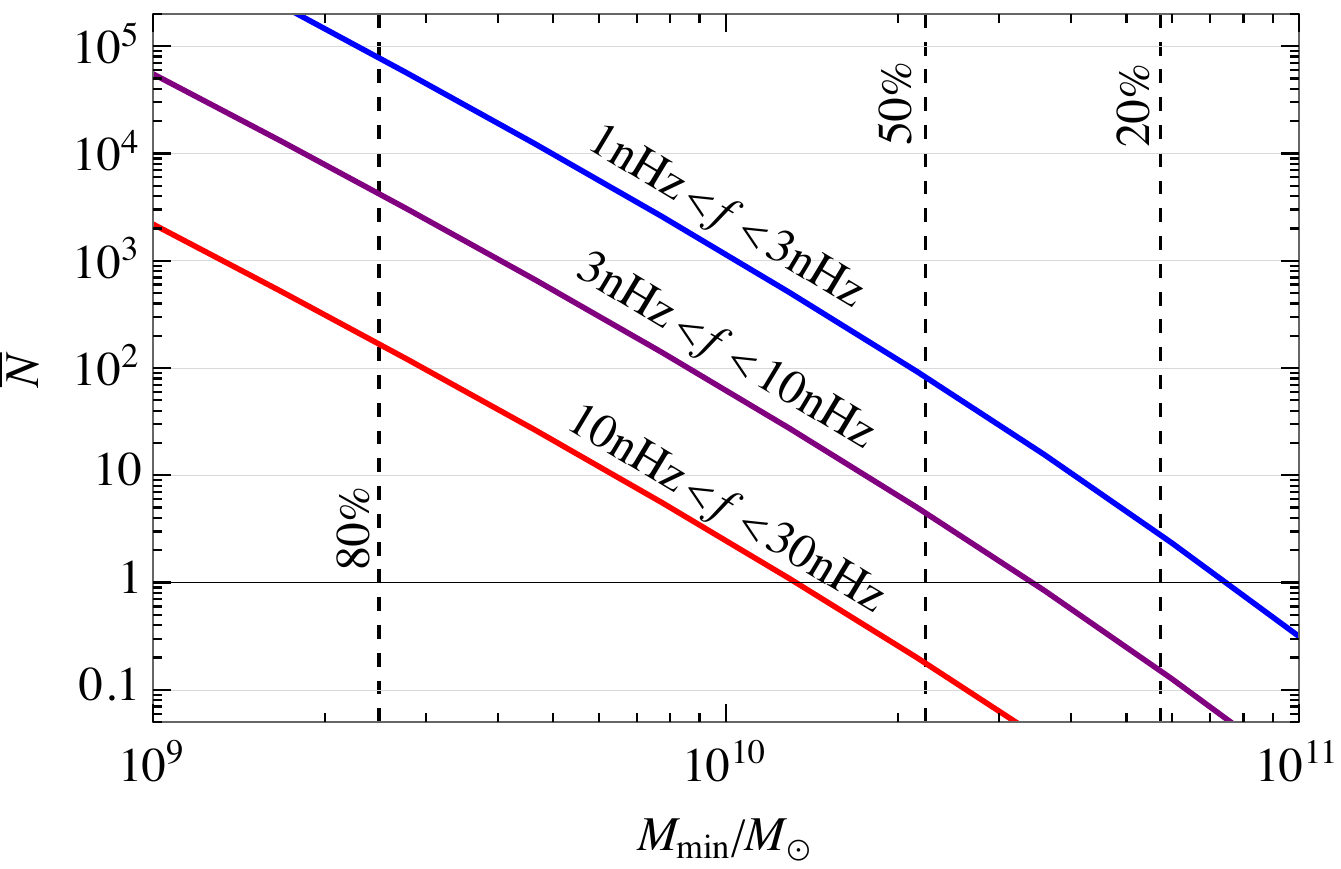}
\caption{The expected number of SMBH binaries heavier than $M_{\rm min}$ and with symmetric mass ratio $\eta>0.01$ emitting GWs in the indicated frequency range for $p_{\rm BH} = 0.17$. The vertical dashed lines indicate the fractions of the total GW signal in each frequency band generated by binaries with larger masses.}
\label{fig:N}
\end{figure}

\noindent
We show in Fig.~\ref{fig:N} the expected numbers of binaries with total masses $M \geq M_{\rm min}$ and mass ratios $q \geq 0.01$ that emit GWs in several frequency bands in the range $1\,{\rm nHz} < f < 30\,{\rm nHz}$ (corresponding to the sensitivity ranges of PTAs and SKA). As seen in Fig.~\ref{fig:SGWB}, SMBH binaries generate the dominant contribution to the GW signal in this frequency range. The fractions of the total GW signal in each frequency band that are generated by binaries with larger masses are indicated by the vertical dashed lines. For example, we see that almost $10^5$ ($10^2$) binaries with $M > 3 \times 10^9 \Msun$ generate 80\% of the calculated GW signal in the $1\,{\rm nHz} < f < 3\,{\rm nHz}$ ($10\,{\rm nHz} < f < 30\,{\rm nHz}$) frequency band, whereas 50\% of the calculated GW signal in the $3\,{\rm nHz} < f < 10\,{\rm nHz}$ band may be generated by just one source. These examples indicate that the expected signal, particularly in higher-frequency bins, may be comprised of a limited number of nearly monochromatic signals from heavy $>10^9 \Msun$ SMBH binaries, 
%with the closest expected around $z \approx 0.2$,
implying sizeable fluctuations around the smooth $f^{\frac23}$ spectrum that would be obtained in the limit of a large population of heavy inspiralling binaries.

\begin{figure*}
\centering
\includegraphics[width=0.85\textwidth]{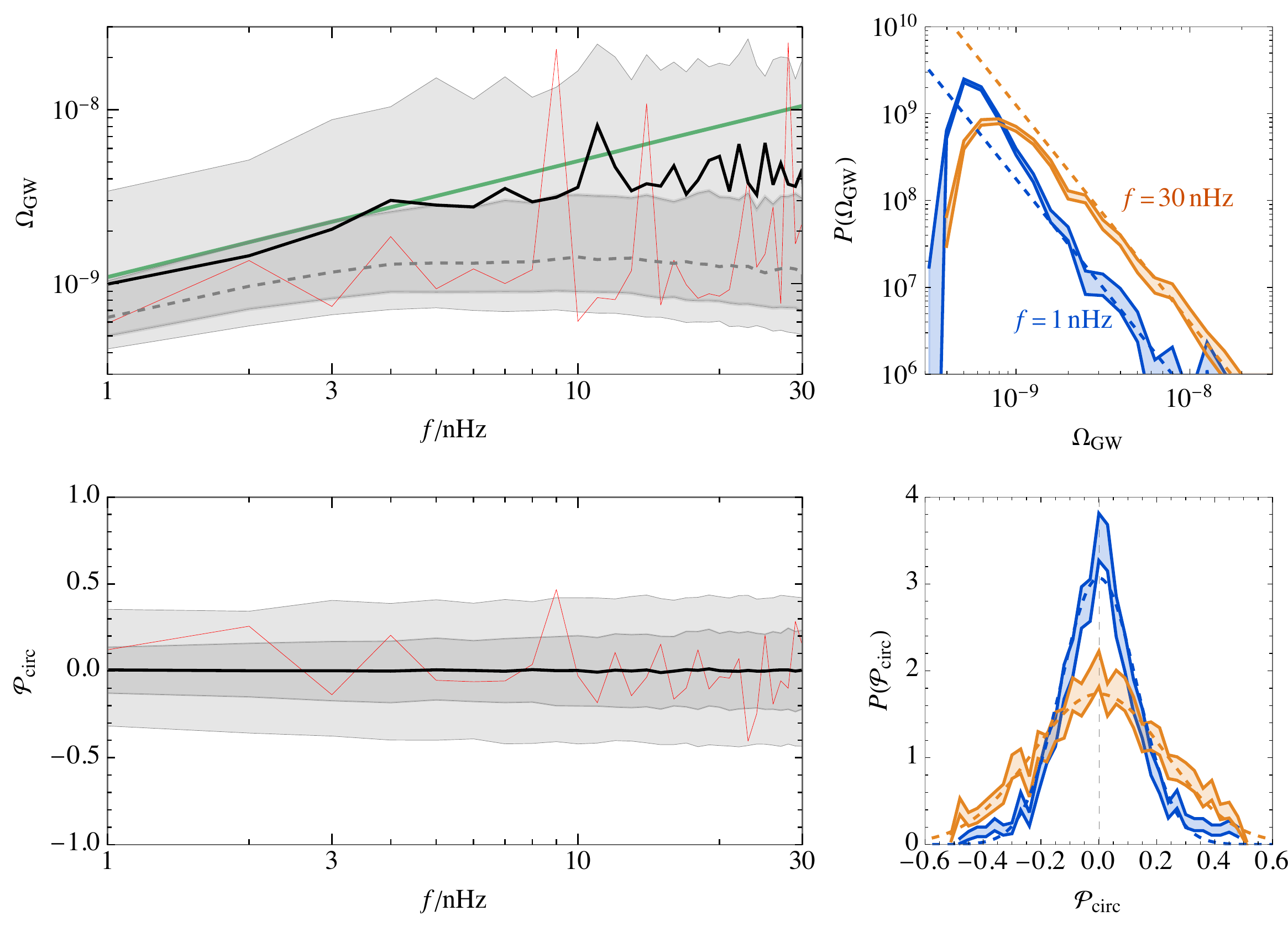}
\caption{The GW energy spectrum $\Omega_{\rm GW}$ and the fractional circular polarization $\mathcal{P}_{\rm circ}$ calculated from 1600 Monte Carlo realisations of the SMBH binary population. In the left panels, the black solid curve shows the mean and the gray bands the $68\%$ and $95\%$ CL regions of the Monte Carlo realisations, and the red curve shows one of the realizations. In the upper left panel the green line shows the mean GW energy density~\eqref{eq:OmegaGW}. The distributions of $\Omega_{\rm GW}$ and $\mathcal{P}_{\rm GW}$ at 1\,nHz (blue) and at 30\,nHz (orange) calculated from the Monte Carlo realisations are shown in the right panels. In the upper right panel, the dashed lines show $P \propto \Omega^{-\frac52}$, and in the lower right panel, the dashed curves show Gaussian fits to these distributions of width $\sigma = 0.13$ (blue) and $\sigma = 0.23$ (orange). In this figure, $p_{\rm BH} = 0.17$ was assumed.}
\label{fig:MCGW}
\end{figure*}

In order to study the prospective SMBH binary population in the PTA band via a Monte Carlo approach, we generate realizations of the population from the probability density function of binaries emitting at a given frequency. We divide the frequency range into bins $(f_j,f_{j+1})$ populated with $N(f_j)$ binaries. A realization of the GW background from each bin can then be obtained as (see Eq.~\eqref{eq:Omega_GW_j})
\be
    \Omega_{\rm GW}(f_j) 
    %= \sum_{k=1}^{N(f_j)} \frac{f}{\Delta f} \left| \frac{\td \tau}{\td \ln f} \right|^{-1} \frac{2\pi}{5} \frac{f^3 |\tilde h(f)|^2}{\rho_c} \bigg|_{\vec{\theta} = \vec{\theta}_k}\, ,
    = \frac{1}{\ln(f_{j+1}/f_j)}\sum_{k=1}^{N(f_j)} \Omega_{\rm GW}^{(1)}(\vec{\theta}_k)  \,,
\ee
where the contribution from an individual binary emitting in this frequency band is given by
\be
    \Omega_{\rm GW}^{(1)}(\vec{\theta}) 
    = \left| \frac{\td \tau}{\td \ln f} \right|^{-1} \frac{2\pi}{5} \frac{f^3 |\tilde h(f)|^2}{\rho_c} \,,
\ee
and $\vec{\theta} \equiv \{\mathcal{M}_z,z,\eta,f\}$ are parameters describing the binary. The number of binaries $N(f_j)$ in this frequency bin is drawn from a Poisson distribution with the expectation value $\bar N(f_j) \equiv \int_{f_{j}}^{f_{j+1}} \!\td N$ 
%\be \label{eq:Nf}
%   \bar N(f_j)
%   = \int_{f_{j}}^{f_{j+1}} \!\td f \int \td \lambda \,\left|\frac{\td \tau}{\td f}\right| \,,
%   = \int_{f_{j}}^{f_{j+1}} \!\td \ln f \, n(f) \, .
%\ee 
and the binary parameters are generated randomly according to the distribution 
\be 
    p(\mathcal{M}_z, \eta, z, f) \propto \frac{\td N}{\td \mathcal{M}_z \td \eta \td z \td \ln f} \, .
\ee
This is the expression that was used to calculate the distributions of binary parameters shown in Fig.~\ref{fig:distr} for a sampling of $10^6$ SMBH binaries with $p_{\rm BH} = 0.17$, redshifts $0<z<3$, mass ratios $\eta > 0.01$ and GW frequencies $1\,{\rm nHz} < f < 30 {\rm nHz}$. We display only binaries with chirp mass $\mathcal{M}_0 > 3\times 10^8 (f/{\rm nHz})^{-\frac43}$, so that $\bar N(f_j) > \mathcal{O}(10^4)$ for each $1\,{\rm nHz}$ frequency bin.

In order to generate the statistical distributions of $\Omega_{\rm GW}$, it is sufficient to consider a smaller set of parameters, and the distribution \eqref{eq:N} for the full population  can be reduced to a simpler distribution $P(\Omega^{(1)},f)$ for individual sources $\Omega_{\rm GW}^{(1)}$ emitting at frequency $f$, which can be expressed via two one-parameter functions depending on the merger rate model, as discussed in Appendix \ref{app:A}.

The upper left panel of Fig.~\ref{fig:MCGW} illustrates the frequency spectra found in 1600 Monte Carlo realizations of the SMBH binary population drawn from a statistical distribution similar to that shown in Fig.~\ref{fig:distr}. The black solid curve shows the mean of the spectra, the grey bands show the $1\sigma$ and $2\sigma$ CL regions of the GW spectra, and the red line is the spectrum found in one representative realization. We note the importance of fluctuations, particularly at higher frequencies where fewer SMBH binaries contribute, with individual binaries becoming distinguishable at frequencies $f \gtrsim 10$~nHz. We find that the median spectrum, shown by the dashed line, has a somewhat lower slope than the analytic result shown as the green line. Although the numerically obtained mean spectrum (black), lies below the analytic expectation, it will approach it slowly if the number of Monte Carlo realizations is increased. The tendency of the mean spectrum to lie along the upper side of the $1\sigma$ CL range is due to the fact that the spectrum has a long tail at high values of $\Omega_{\rm GW}$ generated by occasional nearby binaries: the median spectrum always lies within the $1\sigma$ CL range. All in all, consistently with earlier Monte Carlo studies of the GW signal~\cite{Sesana:2008mz,2011MNRAS.411.1467K}, we find that typical spectra tend to fall below the  $\Omega_{\rm GW} \propto f^{\frac23}$ expectation in higher frequency bins, while a few bins display sharp peaks. 

As there are a finite number of sources, Eq.~\eqref{eq:OmegaGW} is subject to statistical fluctuations, which arise mostly from the possibility of having a few strong sources nearby. To obtain an order-of-magnitude estimate, we focus on the closest binaries and ignore the redshift dependence. In this case,
$\Omega^{(1)}_{\rm GW} \propto D_{L}^{-2}$, by  Eqs.~\eqref{eq:A(f)} and \eqref{eq:OmegaGW}, and the probability to find an event at $D_{L}$ is $P^{(1)}(D_{L}) \propto D_{L}^2$. Thus 
$\langle \Omega^{(1)}_{\rm GW} \rangle \propto \int^{D_{L, \rm max}}_{D_{L, \rm min}} \td D_{L} P(D_{L}) \Omega^{(1)}_{\rm GW} \propto D_{L, \rm max}$, 
where, in this simplified approach, $D_{L, \rm max}$ is some large luminosity distance at which $\Omega^{(1)}_{\rm GW}$ gets suppressed, and $D_{L, \rm min}$ is the distance to the nearest possible source. On the other hand, $\langle (\Omega^{(1)}_{\rm GW})^{2} \rangle \propto 1/D_{L, \rm min}$. Therefore, we expect the mean of the GW signal to be determined by faraway sources, while the variance is set by a few close-by binaries. Moreover, we can estimate from $P^{(1)}(D_{L})$ that $\Omega^{(1)}_{\rm GW}$ has a relatively flat power-law tail at large values, 
\be\label{eq:P_Omega_tail}
    P^{(1)}(\Omega) \propto \Omega^{-\frac52}, \quad \mbox{when} \quad
    \Omega \to \infty \,.
\ee
Since the closest distance to massive BH binaries is constrained, this tail will be cut off at $\Omega^{(1)}_{\rm GW, max} \propto D_{L, \rm min}^{-2}$. Even with this cutoff, the mean and the variance are still not very useful characteristics of the uncertainties, and we find it more illuminating to estimate the confidence intervals around the median value.

The upper right panel of Fig.~\ref{fig:MCGW} displays the distributions of $\Omega_{\rm GW}$ at 1\,nHz (blue) and at 30\,nHz (orange) found in 1600 Monte Carlo realisations of the SMBH binary population. We note that the distributions at both frequencies have tails that approach the analytic result  $P \propto \Omega^{-\frac52}$ (\ref{eq:P_Omega_tail}), as indicated by the dashed lines. These tails exhibit explicitly why the mean (black solid) line in Fig.~\ref{fig:MCGW} tends to lie above  the 68\% CL band, while the median (gray dashed) line lies within it at all frequencies. We note that the overall shapes of the distributions resemble the analytic calculations shown in the left panel of Fig.~\ref{fig:naive} in Appendix~\ref{app:A}. In summary, even if a bin has a high number of contributing events (i.e., $\bar N \to \infty$), the distribution of $\Omega_{\rm GW}$ does not converge to a Gaussian because it retains its $\Omega_{\rm GW}^{-5/2}$ tail. Since the variance is not well behaved, we define the width of the distribution $\Delta_{\Omega_{\rm GW}}$ as the width of the 68\% confidence interval, as in Fig.~\ref{fig:MCGW}. The width-to-mean ratio scales as
\be
    \Delta_{\Omega_{\rm GW}}/\left\langle\Omega_{\rm GW}\right\rangle 
    \propto \bar N^{-\frac13} \, ,
\ee
when $\bar N \gg 1$, as is shown in Appendix~\ref{app:A}. This scaling is slower than the typical $1/\sqrt{\bar N}$ scaling predicted by the central limit theorem.

\subsection{GW polarization}
\label{sec:polarization}

Recent studies have argued that the circular polarization of the signal can be used to estimate whether the SGWB comes from a handful of sources or a relatively large population of binaries~\cite{Kato:2015bye, Conneely:2018wis, Hotinli:2019tpc, Belgacem:2020nda, Sato-Polito:2021efu,ValbusaDallArmi:2023ydl}. We recall that the left and right circular GW polarization amplitudes from a binary with inclination angle $\theta$ are
\be\label{eq:hL,R}
    |h_{L,R}(f)|^2 = \frac{1}{16}(1 \pm \cos{\theta})^4 \, |h(f)|^2 \,,
\ee
where 
\be \label{eq:h_t_inspiral}
    |h(f)| 
    = \frac{4 \mathcal{M}_z^{\frac53}}{D_L} (\pi f)^{\frac{2}{3}}
\ee
is the maximal GW strain from an inspiralling binary, and the gravitational Stokes parameters are defined by
\bea\label{eq:I,V}
    &I(f) = |h_L(f)|^2 + |h_R(f)|^2 \,, \\
    &V(f) = |h_L(f)|^2 - |h_R(f)|^2 \,.
\eea
The fractional amount of circular polarization of the SGWB can be characterized by the quantity
\be\label{eq:Pcirc}
    \mathcal{P}_{\rm circ}(f) = \frac{\sum_i V_i(f)}{\sum_i I_i(f)}\,,
\ee
where the sums are over all binaries in a fixed frequency range. 

If the GW signal is dominated by a single source, then $\mathcal{P}_{\rm circ}$ depends only on the inclination angle and large circular polarizations are preferred, with $\mathcal{P}_{\rm circ} > 0.87(0.2)$ at the 68\% CL (95\% CL). On the other hand, if the GW signal is dominated by several ($N_{\rm dom} \gtrsim 10$) sources of comparable strengths, then the $\mathcal{P}_{\rm circ}(f)$ will be approximately Gaussian with a width determined by fluctuations in $\sigma_{V}$. Since $\sigma_{V}/\langle I\rangle_{\theta} = 1.17$ for a single source, we can estimate that (see Appendix \ref{app:A} for details)
\be\label{eq:sigma_circ}
    \sigma_{\mathcal{P}_{\rm circ}} 
    \approx \sigma_{V}/\langle I\rangle_{\theta} 
    \approx 1.17/\sqrt{N_{\rm dom}} \,.
\ee

The lower left panel of Fig.~\ref{fig:MCGW} illustrates the distributions of the circular polarization  $\mathcal{P}_{\rm circ}(f)$ found in the sample of 1600 Monte Carlo realizations of the SMBH binary population with $M > 10^9\Msun$ and $\eta > 0.01$ whose frequency spectra were illustrated in Fig.~\ref{fig:MCGW}. We see that the mean value of $\mathcal{P}_{\rm circ}(f) \approx 0$, as expected, but large statistical fluctuations are possible even at the $1\sigma$ level. This phenomenon is visible in the red curve, which shows results from one of the Monte Carlo realizations. The large fluctuations reflect the fact that the observable GW spectrum could be due to a very limited number of sources,
particularly at higher frequencies.

The lower right panel of Fig.~\ref{fig:MCGW} displays the distributions of $\mathcal{P}_{\rm circ}$ at 1\,nHz (blue) and at 30\,nHz (orange) found in 1600 Monte Carlo realizations of the SMBH binary population. We note that the statistical distribution is indeed broader at the higher frequency, as expected. The dashed curves are Gaussian fits to polarization distributions with widths $\sigma = 0.13$ (blue) and $\sigma = 0.23$ (orange). These fits are very accurate: see also the right panel of Fig.~\ref{fig:naive} in Appendix~\ref{app:A}.

As the variance of the circular polarization of the signal depends on the number of binaries that dominate the signal, it would be suppressed if the merger rate of the heaviest binaries were suppressed, which could be the case if $p_{\rm BH}$ is not universal. By suppressing $p_{\rm BH}$ e.g. at $M>10^9 M_\odot$ and enhancing $p_{\rm BH}$ for $10^9 M_\odot > M>10^6 M_\odot$ one accommodate the IPTA common-spectrum effect. In this case, the GW energy spectrum $\Omega_{\rm GW}$ becomes smoother and approaches the naive $\Omega_{\rm GW} \propto f^{\frac23}$ behavior while suppressing the circular polarization $\mathcal{P}_{\rm circ}$. Moreover, in this case, $p_{\rm BH}$ for the lighter binaries would be enhanced, which would increase the number of signals in the sensitivity range of LISA and possibly AEDGE.

\subsection{Prospects for future GW observatories}
\label{sec:prospects}

\begin{figure*}
\centering
\includegraphics[width=1.5\columnwidth]{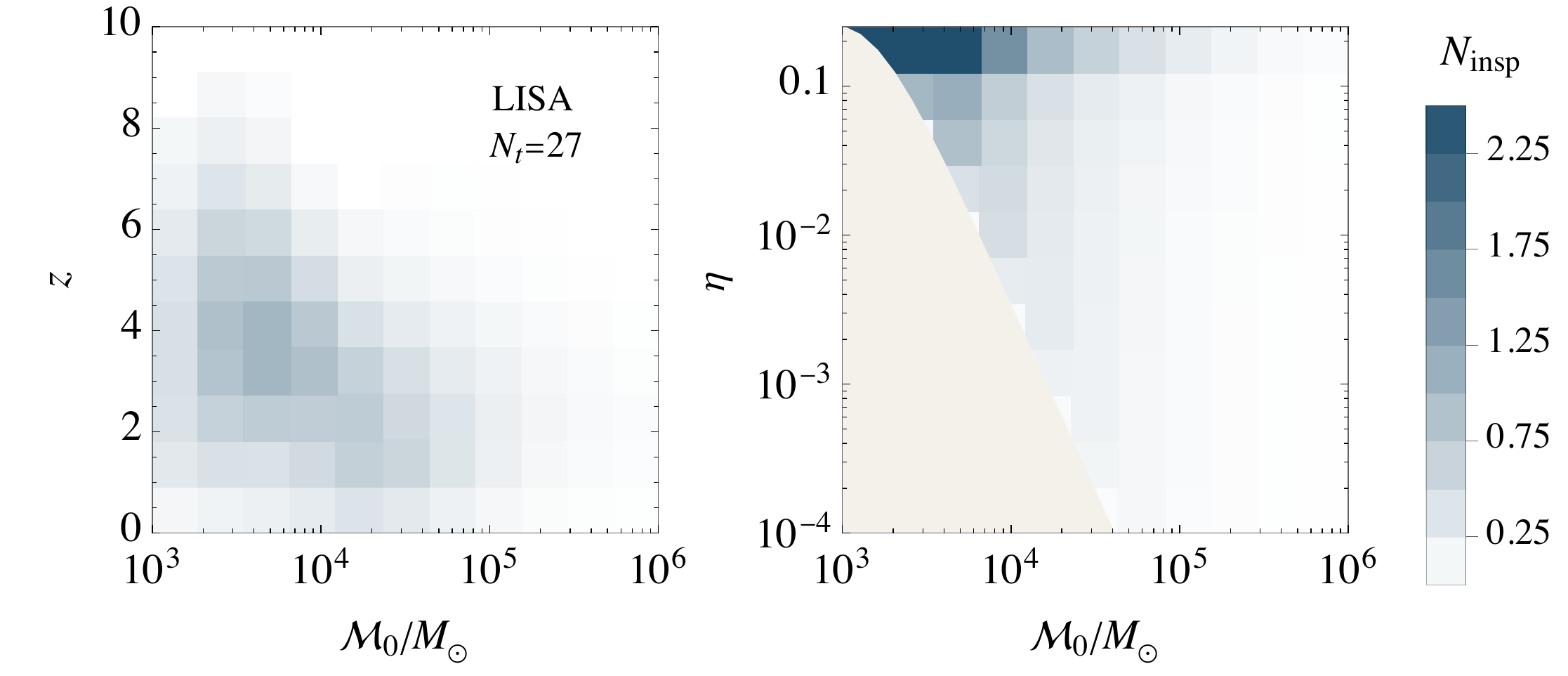}
\caption{The expected numbers of near-monochromatic GW signals generated more than a day before the merger that would be detectable in a year of LISA observation, as functions of the chirp mass $\mathcal{M}_0$ and redshift $z$ (left panel) and of the chirp mass and symmetric mass ratio $\eta$ (right panel). The merger rates assume the abundance of IMBHs described by Eq.~\eqref{mbhmh} for BHs heavier than $10^3 M_{\odot}$ and a universal merger efficiency factor of $p_{\rm BH}=0.17$ inferred from IPTA data, as discussed in the text.}
\label{fig:insp_prosp}
\end{figure*}

As discussed earlier, BH binaries can generate two qualitatively distinct classes of signal: long, nearly monochromatic signals from the slowly-evolving inspiralling phase and relatively short signals from the merger and ringdown. The detections of these two types of sources need to be considered separately. 
 
\begin{figure*}
\centering
\includegraphics[width=1.5\columnwidth]{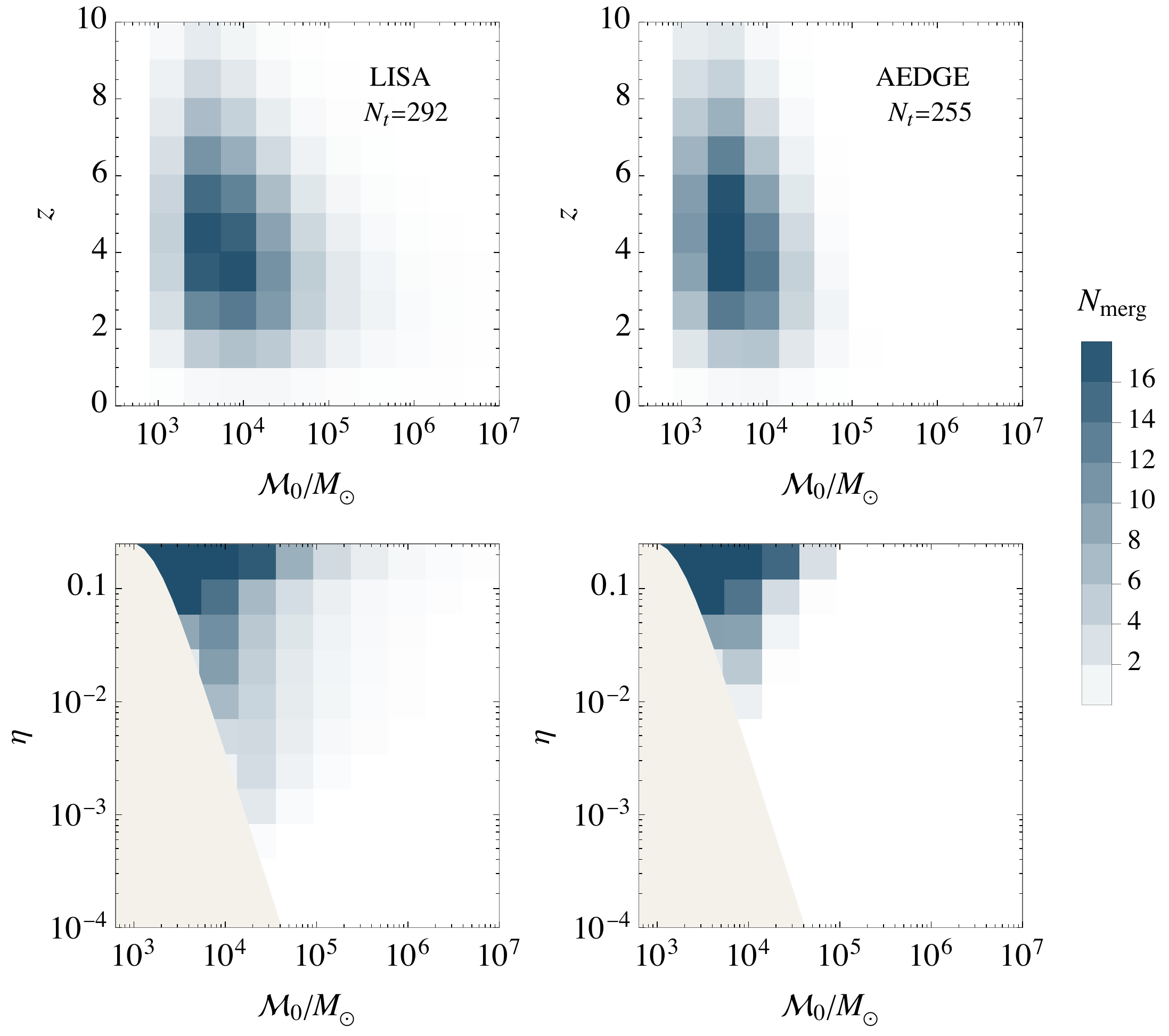}
\caption{The expected numbers of detectable GW events generated during the last day before the merger in a year of observation by LISA (left panels) and AEDGE (right panels) as functions of the chirp mass $\mathcal{M}_0$ and redshift $z$ (upper panels) and of chirp mass and symmetric mass ratio $\eta$ (lower panels). The merger rates assume the abundance of IMBHs described by Eq.~\eqref{mbhmh} for BHs heavier than $10^3 M_{\odot}$ and a universal merger efficiency factor of $p_{\rm BH}=0.17$ inferred from IPTA data.}
\label{fig:prospects}
\end{figure*}

The number of detectable nearly monochromatic GW signals that arise from inspiralling BH binaries is
\be
    N_{\rm insp} 
    = \int_{\mathcal{T}+\tau_{\rm min}}^\infty \!\!\td \tau \int \td \lambda\,p_{\rm det}\!\left[\frac{{\rm SNR}_c}{{\rm SNR}(\tau,z)}\right] \, ,
\ee
where we account only for binaries whose coalescence time $\tau$ is longer than $\tau_{\rm min} = 1\,$day and we estimate the signal-to-noise ratio (SNR) for a detector characterized by the noise power spectrum $S_n(f)$ as
\be
    {\rm SNR}(\tau,z) = \sqrt{\int_0^\mathcal{T} \td t \,\frac{2 |h(\tau-t,z)|^2}{S_n(f(\tau-t,z))}} \,.
\ee 
The optimal time-dependent inspiral strain $|h(t)| = |h(f(t))|$ is given by Eq.~\eqref{eq:h_t_inspiral} and the GW frequency as a function of time is given by Eq.~\eqref{eq:tau(f)}. Analogously, the expected number of BH binary merger events is
\be
    N_{\rm merg} =  \mathcal{T} \int \td \lambda \,p_{\rm det}\!\left[\frac{{\rm SNR}_c}{{\rm SNR}(z)}\right] \, ,
\ee
where
\be
    {\rm SNR}(z) = \sqrt{\int_{f(\tau_{\rm min})}^\infty \td f \,\frac{4 |\tilde h(f)|^2}{S_n(f)}} 
\ee 
The cut at $f(\tau_{\rm min})$ here implies that we consider the SNR only from the last day of the signal. In both cases, the detection probability $p_{\rm det}$ accounts for the detector's antenna patterns and includes the average over the binary inclination, sky location, and polarization~\cite{Finn:1992xs, Gerosa:2019dbe}, the noises $S_n$ include the foregrounds from stellar mass BH binaries and white dwarf binaries (see e.g.~\cite{Lewicki:2021kmu}), and we use ${\rm SNR}_c = 8$ for the detection threshold and $\mathcal{T} = 1\,{\rm year}$ for the observation time.

We find that LISA is the only detector capable of measuring the near-monochromatic sources arising in the early stages of the inspiral phase, i.e., more than 1 day from the beginning of the merger phase. AEDGE cannot see these near-monochromatic sources because the coalescence time of binaries heavier than $10^3\Msun$ is less than a day when they enter the AEDGE sensitivity window.\footnote{In this work we have cut the AEDGE sensitivity at $3$\,mHz. This cut is motivated by the potential Newtonian gravity backgrounds~\cite{Hogan:2011tsw,Graham:2017pmn}. We note, however, that dedicated studies of these backgrounds for AEDGE have not been performed.} Fig.~\ref{fig:insp_prosp} illustrates the prospects to detect the inspiral signals with LISA. The shades in the panels correspond to the number of events whose parameters fall within each rectangular bin. The distribution peaks below $\mathcal{M}_0 = 10^4 M_{\odot}$ reflecting the binary population, that increases towards lower masses, and the cut-off at $10^3\Msun$. Similarly, because the GW signal \eqref{eq:A(f)} from inspirals does not explicitly depend on $\eta$, the $\eta$ distribution in the right panel of Fig.~\ref{fig:insp_prosp} arises solely from the binary population. The binary merger rate peaks at $4<z<6$ (see Fig.~\ref{fig:RBH}) but LISA can not see the majority of the binaries beyond $z\sim 4$.

We find that, at $z<10$, there are a total of $\mathcal{O}(10^7)$ near-monochromatic sources in the frequency range $10^{-5} {\rm Hz} < f < 0.1{\rm Hz}$.~\footnote{We emphasize, however, that there could be lower-mass BHs that are remnants of Population III stars, which are not included in our analysis.} As only a small fraction of these can be resolved by LISA, we expect that the rest constitute a significant stochastic GW background. We leave a detailed study of this background and its detectability with LISA for future work.

Fig.~\ref{fig:prospects} illustrates AEDGE and LISA prospects for observing GW events from BHs less than a day from the merger. The expected total number of detectable events is slightly larger for LISA than for AEDGE. Due to the different frequency ranges, LISA can see heavier binaries, $\mathcal{M}_0>10^4\Msun$, while AEDGE can see lighter binaries, $\mathcal{M}_0<2\times10^3\Msun$. LISA could in principle detect even heavier mergers, $\mathcal{M}_0\lesssim 10^8\Msun$, but since such mergers are so rare it is unlikely that LISA will see any mergers above $10^6\Msun$. The expected number of detectable mergers peaks at $4<z<6$ for both of the experiments. For AEDGE, the $z$ distribution of the  detectable mergers reflects the binary merger rate (cf. Fig.~\ref{fig:RBH}) while for LISA the $z$ distribution peaks at slightly lower $z$ because the SNR of $\mathcal{M}_0\sim 10^4$ binaries for LISA is not as high as it is for AEDGE. The lower panels show that, for all detectors, the BH mass ratio peaks at the highest values. For AION-km and ET we find that the total expected number of events per year is less than three.

We note that most of the IMBH events detectable by AEDGE during the last day prior to the merger (see the right panels in Fig.~\ref{fig:prospects}) will also have been detected by LISA during the previous infall stage, as seen in Fig.~\ref{fig:insp_prosp}. This opens prospects for using LISA data to predict when and in what direction AEDGE will observe IMBH mergers, sharpening tests of general relativity and giving advance warnings for searches for possible multi-messenger signals.

We caution that there are observational uncertainties in the low mass cut-off in Eq.~\eqref{mbhmh}, due to the difficulty of measuring very faint active galactic nuclei (AGNs) or inactive BHs in dwarf galaxies. As mentioned above, surveys such as eRASS and AMUSE~\cite{Miller:2014vta} are already constraining this low-mass region and are compatible with the assumed cut-off, although a heavier mass cut-off may be favored~\cite{Chadayammuri:2022bjj}. This could easily be achieved in models with modified initial fluctuation spectra~\cite{Hutsi:2022fzw}. 

The cut-off mass is tightly related to the SMBH formation mechanism since a BH in the halo centre cannot be lighter than the seed that originated the growth, for a review see~\cite{Volonteri:2021sfo}. In this sense, the cut-off we have considered assumes the existence of some light seeds, which augment the possible GW signal.
Independently of the growth through halo merging, IMBHs can also be born inside dense stellar media like nuclear and globular star clusters. The growth from stellar masses happens because of repeated encounters inside these dense environments. Estimating the GW emission from such encounters is an active field of research~\cite{Fragione:2022ams,Fragione:2018yrb,Fragione:2018vty} and their signals may contribute significantly to the expected number of IMBH mergers.

%%%%%%%%%%%%%%%%%%%%%%%%%%%%%%%%%%%%%%%%%%%
\section{Conclusions}
%%%%%%%%%%%%%%%%%%%%%%%%%%%%%%%%%%%%%%%%%%%

We have described the results from a model for the PTA nHz common-process signal based on a simulation of massive BH mergers. The magnitude of the signal depends on the merger probability, $p_{\rm BH}$, which is a product of the probabilities that a pair of halos contain massive BHs and the probability that they merge. We have sidestepped the considerable uncertainties in modelling these probabilities by fitting a mass-independent value of $p_{\rm BH}$ to the PTA signal, finding  $p_{\rm BH} = 0.17$ with a factor of two uncertainty. With this assumption, the dominant contribution to the PTA signal is made by mergers with total masses $> 10^9 \Msun$ and about 10\% from masses $< 10^9 \Msun$. The PTA mergers would have redshifts ${\cal O}(1)$ and have mass asymmetries $\gtrsim 10$.

The number of mergers contributing most of the PTA signal at frequencies ${\cal O}(10)$~nHz is limited. Consequently, the frequency spectrum becomes quite irregular, the spectral index may deviate from the analytic value of 2/3, individual mergers may be distinguished, and there may be detectable circular polarization. These will be interesting targets for future experiments in the nHz range, including PTAs and SKA.

Assuming the same mass-independent value of  $p_{\rm BH}$ as for the PTA signal, there would be observable signals from mergers with total masses $\in (10^3, 10^6) \Msun$ in the LISA experiment and from mergers with total masses $\in (10^3, 10^5) \Msun$ in the AEDGE experiment. Data from these experiments will be able to check the accuracy of the constancy of $p_{\rm BH}$. In principle, their signals could be even larger than our estimates if $p_{\rm BH}$ is closer to unity for masses below the PTA range, or if there is a significant GW contribution seeded by Population III stars, though a lower value of $p_{\rm BH}$ and hence a lower event rate cannot be excluded.

\vspace{8pt}
\begin{acknowledgments}
\vspace{-9pt}
This work was supported by European Regional Development Fund through the CoE program grant TK133 and by the Estonian Research Council grants PRG803 and PSG869. The work of J.E. was supported by the United Kingdom STFC Grant ST/T000759/1. The work of V.V. was supported by the European Union's Horizon Europe research and innovation programme under the Marie Sk\l{}odowska-Curie grant agreement No. 101065736.
\end{acknowledgments}

\appendix

\section{Statistics of the GW energy spectrum and circular polarization}
\label{app:A}

%\R{[Include plot of $\Omega_{\rm GW}$ probability distribution in some bins from MC and show the -5/2 tail]}

\subsection{Distribution of $\Omega_{\rm GW} (f)$} 

The GW energy spectrum arising from a population of BH binaries can be expressed as the sum of the contributions from individual binaries. Consider first the contribution to $\Omega_{\rm GW}$ coming from a single inspiralling binary,
\be
    \Omega_{\rm GW}^{(1)}(\vec{\theta}) 
    = \left| \frac{\td \tau}{\td \ln f} \right|^{-1} \frac{2\pi}{5} \frac{f^3 |\tilde h(f)|^2}{\rho_c} \, ,
\ee
where $\vec{\theta} \equiv \{\mathcal{M}_z,z,f\}$ are parameters describing the binary. The GW energy spectrum arising from the inspiralling binary population (or any of its subpopulations) arises from the sum of nearly monochromatic components
\be
    \Omega_{\rm GW}(f) \approx \sum_{i} \Omega_{{\rm GW}}^{(1)}(\vec{\theta}_i) \delta(\ln f/f_i) \, .
\ee
This approximation is valid as long as the change in frequency over the observation period is smaller than the spectral resolution. The parameters $\vec{\theta}_i$ as well as $\Omega_{{\rm GW}}^{(1)}$ are independent and identically distributed. Thus, the statistical properties of $\Omega_{\rm GW}(f)$ can be inferred from the distribution of $\Omega_{{\rm GW}}^{(1)}(f)$, which is given by
\bea
    P^{(1)}(\Omega|f) 
    &= \frac{1}{n(f)} \int \td \lambda \left| \frac{\td \tau}{\td \ln f} \right| 
    \delta\left( \Omega - \Omega^{(1)}_{\rm GW} \right) \\
    &= \frac{\mathcal{F}(\Omega f^{-\frac{10}{3}})}{f \Omega^{\frac32} n(f) } \, ,
\eea
where we have defined the function
\bea
    \mathcal{F}(x) 
    \equiv \int \td z & \td \eta \bigg[ \frac{\partial \lambda (\mathcal{M}_z,z)}{\partial \mathcal{M}_z \partial \eta \partial z} \\
    & \ \times \frac{\mathcal{M}_z}{16 \sqrt{10 \rho_c} D_L} \, \bigg]_{\mathcal{M}_z = \frac{1}{\pi} \left(\frac{5\pi}{8}D_L^3 \rho_c x \right)^{\frac{3}{10}}}
%    &\mathcal{F}(x)     \equiv \int \td z \td \eta\frac{\td \lambda (\mathcal{M}_z(z,x),z)}{\td \mathcal{M}_z \td \eta \td z} \frac{\mathcal{M}_z(z,x)}{16 \sqrt{10 \rho_c} D_L} \, |_{\mathcal{M}_z = \frac{1}{\pi} \left(\frac{5\pi}{8}D_L^3 \rho_c x \right)^{\frac{3}{10}}} : \\
%    &\mathcal{M}_z(z,x) 
%    = \frac{1}{\pi} \left(\frac{5\pi}{8}D_L^3 \rho_c x \right)^{\frac{3}{10}} \, ,
\eea
and the spectral source density
\be
    n(f) 
    \equiv \int \td \mathcal{M}_z \td \eta \td z \frac{\td N}{\td \mathcal{M}_z \td \eta \td z \td \ln f} \,.
\ee
We see that, for circular inspiralling binaries, the 4-parameter distribution \eqref{eq:N} can be reduced to two functions of a single parameter. Further reduction is unlikely, as these functions depend on the model of the binary merger rate, which we assume to stem from the merger rate of galaxies.

The average energy spectrum in a frequency bin $(f_j,f_{j+1})$ can be obtained from 
\bea\label{eq:Omega_GW_j}
    \Omega_{{\rm GW}}(f_j) 
    &\equiv \frac{1}{\ln (f_{j+1}/f_j)} \int^{f_{j+1}}_{f_j} \Omega_{{\rm GW}}(\vec{\theta};f) \td \ln f\\ 
    &= \frac{1}{\ln (f_{j+1}/f_j)} \sum^{N(f_j)}_{i=1} \Omega^{(1)}_{{\rm GW},i} \, ,
\eea
where the number of binaries $N(f_j)$ is drawn from a Poisson distribution with the expected value,
\be 
   \bar N(f_j)
%   = \int_{f_{j}}^{f_{j+1}} \!\td f \,\left|\frac{\td \tau}{\td f}\right| \int \td \lambda \,.
   = \int_{f_{j}}^{f_{j+1}} \!\td \ln f \, n(f) \, ,
\ee 
and the $\Omega^{(1)}_{{\rm GW},i}$ are drawn from a random distribution $P^{(1)}(\Omega,f) = P^{(1)}(\Omega|f) n(f)/\bar N(f_j)$ with $f \in (f_j,f_{j+1})$. In the following, we will suppress the $f_j$.

The moment generating function of $\Omega_{{\rm GW}}$ is given by
\bea\label{eq:G_1}
    M_{\Omega_{\rm GW}}(s) 
    &\equiv \langle \exp\left(s \Omega_{\rm GW} \right) \rangle \\
    &= \sum_{N \geq 0} p_N \prod^{N}_{i=1} \left\langle \exp\left(s\Omega^{(1)}_{\rm GW} \right) \right\rangle \\
    &= \sum_{N \geq 0} \frac{{\bar N}^{N}}{N!} e^{-\bar N}\left( M_{\Omega^{(1)}_{\rm GW}}(s) \right)^{\bar N} \\
    &= \exp\left[ \bar N \left(M_{\Omega^{(1)}_{\rm GW}}(k) - 1 \right) \right] \, .
\eea
Thus the cumulant generating function of $\Omega_{{\rm GW}}(f_j)$, that is,
\bea
    K_{\Omega_{\rm GW}}(s) 
    &\equiv \ln M_{\Omega_{\rm GW}}(s) \, \\
    &= \int_{f_{j}}^{f_{j+1}} \td N \left[ \exp(s \,\Omega^{(1)}_{\rm GW}) - 1 \right] \, .
\eea
is proportional to the generating function of $M_{\Omega^{(1)}_{\rm GW}}(s)$, implying that the $n$th cumulant of $\Omega_{{\rm GW}}$ is
\bea
    \kappa_{n} [\Omega_{\rm GW}] 
    = \bar N \langle (\Omega^{(1)}_{\rm GW})^{n} \rangle \, 
    = \int_{f_{j}}^{f_{j+1}} \td N (\Omega^{(1)}_{\rm GW})^{n} \, .
\eea
Therefore, the mean $\kappa_{1} (\Omega_{\rm GW}) \equiv \langle \Omega_{\rm GW} \rangle$
matches the expectation value given in Eq.~\eqref{eq:OmegaGW}, while the variance 
\be\label{eq:deltaOmega}
    \kappa_{2} (\Omega_{\rm GW}) 
    \equiv \langle (\delta\Omega_{\rm GW})^{2} \rangle
    = \bar N \langle (\Omega^{(1)}_{\rm GW})^{2} \rangle
\ee 
is divergent due to the long $P^{(1)}(\Omega) \propto \Omega^{-\frac52}$ tail \eqref{eq:P_Omega_tail} unless a minimal distance to the closest BH binary is imposed. In particular, as we will demonstrate briefly, such long tail is preserved in the distribution of $\Omega_{{\rm GW}}$, i.e., the higher cumulants are not diminished when summing several instances of $P^{(1)}(\Omega) \propto \Omega^{-\frac52}$.

The moment generating function $M_{\Omega_{\rm GW}}(-s)$ is a Laplace transform of the probability distribution. Thus as an alternative to the Monte Carlo approach adopted in the main text, the probability distribution of $\Omega_{\rm GW}(f_j)$ can be obtained by an inverse Laplace transform of the moment generating function given in Eq.~\eqref{eq:G_1}. It can be expressed as
\bea\label{eq:P_analytic}
    P(\Omega_{\rm GW}) 
    &= \frac{1}{2\pi \bar N} \int^{\infty}_{-\infty} \td s \,  
    e^{i s \Omega_{\rm GW}/\bar{N} } \times \\
    &\times \exp\left( \bar N \left\langle \exp\left(-i s \Omega^{(1)}_{\rm GW}/\bar N \right) - 1 \right\rangle \right) \, ,
\eea
where the average is taken over the single event distribution $P^{(1)}$.

\paragraph{The large $N$ limit. --} The central limit theorem (and Eq.~\eqref{eq:deltaOmega}) dictates that when $N\to \infty$, then the distribution \eqref{eq:Omega_GW_j} should approach a Gaussian with its relative width scaling as $\sigma_{\Omega_{\rm GW}}/\langle \Omega_{\rm GW} \rangle \propto 1/\sqrt{N}$. However, this is not the case when the second cumulant diverges.  Consider a generic distribution with a tail (we will drop "GW" subindex for the sake of brevity)
\be\label{eq:P1_asympt}
    P^{(1)} (\Omega) \stackrel{\Omega \to \infty}{\sim} C \Omega^{-\frac52}, 
\ee
where $C$ is a constant. The second moment $\langle \Omega^2\rangle$ diverges. We assume that the small $\Omega$ behavior is such that the first moment $\bar \Omega^{(1)} \equiv \langle \Omega^{(1)}\rangle$ is finite. The $\bar N \to \infty$ asymptotic of the average in Eq.~\eqref{eq:P_analytic} can then be computed by separating it into the mean and a term for which the expectation value can be determined by approximating the distribution by its tail \eqref{eq:P1_asympt}:
\begin{align}
    &\bar N \left\langle \exp\left(-i s \frac{\Omega^{(1)}}{\bar N} \right) - 1 \right\rangle \nonumber\\
    =& -i s \bar \Omega^{(1)}  + \bar N \left\langle \exp\left(-i s \frac{\Omega^{(1)}}{\bar N}  \right) - 1 + i s \frac{\Omega^{(1)}}{\bar N} \right\rangle \nonumber\\
    %\stackrel{N \to \infty}{\sim} & -i s \bar \Omega^{(1)}- i C \bar N (s/\bar N)^{\alpha-1} e^{i \frac{\pi}{2} \alpha} \Gamma(1-\alpha) \, . \nonumber
    \stackrel{N \to \infty}{\sim} & -i s \bar \Omega^{(1)} + e^{i \pi \frac34} \frac{\sqrt{\pi}4C}{3\sqrt{\bar N}} s^{\frac32} \, . 
\end{align}
In summary, the $N \to \infty$ distribution asymptotes to
\bea\label{eq:P1_asympt2}
    P(\Omega) 
%    &\approx \frac{1}{\pi \bar{N}} {\rm Re} \int^{\infty}_{0} \td s \, \exp\left(i s (\Omega/\bar{N} - \bar{\Omega}^{(1)}) + e^{i \pi \frac34} \frac{\sqrt{2\pi}4C}{3\sqrt{\bar N}} s^{\frac32} \right) \, , \nonumber\\
%    &\approx \frac{1}{\pi (C\bar{N})^{\frac{2}{3}} } {\rm Re} \int^{\infty}_{0} \td s \, \exp\left(i s (C\bar{N})^{-\frac{2}{3}} (\Omega - \bar{N}\bar{\Omega}^{(1)}) + e^{i \pi \frac34} \frac{\sqrt{2\pi}4}{3} s^{\frac32} \right) \, , \nonumber\\
    &\approx (C\bar{N})^{-\frac{2}{3}} \mathcal{P}\left( (C\bar{N})^{-\frac{2}{3}} ( \Omega - \bar \Omega )\right) \, ,
\eea
where $\bar{\Omega} \equiv \bar{N}\bar{\Omega}^{(1)}$ denotes the expectation value of $\Omega$ and
\bea
    \mathcal{P}\left( x \right) \equiv 
    %{\rm Re} \int^{\infty}_{0} \frac{\td s}{\pi} \, \exp\left(i x s + e^{i \pi \frac34} \frac{\sqrt{\pi}4C}{3} s^{\frac32} \right) \, 
    {\rm Re} \frac{e^{i\pi/6}}{\pi} \int^{\infty}_{0} \td s \, \exp\left(i e^{i\pi/6} x s - \frac{4\sqrt{\pi}}{3} s^{\frac32} \right) \, 
\eea
is a universal function that is independent of the single event distribution -- all information about the latter enters through $\bar{\Omega}^{(1)}$ and $C$. This limiting case reveals a few crucial features: 
\begin{enumerate}[leftmargin=*]
    \item Since $\mathcal{P}(x) \sim x^{-5/2}$ when $x \to \infty$, the power-law behavior of the tail is preserved in the large $\bar N$ limit:
\be\label{eq:52tail}
    P(\Omega) \stackrel{\Omega \gg \bar{\Omega}}{\approx} \frac{C\bar{N}}{(\Omega- \bar{\Omega})^{\frac52}}\, .
\ee
    \item When the number of contributing events $N$ is increased by a factor of $A$ to $AN$, the distribution evolves as
\be\label{eq:52scaling}
    P^{(AN)}(\Omega)\!=\!
    A^{-\frac23} P^{(N)}\left(\!A^{-\frac23}(\Omega\!-\! \bar{\Omega}^{(AN)})\!+\!\bar{\Omega}^{(N)}\!\right).
\ee
    \item Although the variance diverges, we can define the width of the distribution $\Delta_{\Omega} \equiv \Omega_+ - \Omega_-$ from the confidence interval $(\Omega_-,\Omega_+)$ centred around, e.g., the median. Eq.~\eqref{eq:52scaling} then implies the scaling
\be
    \Delta_{\Omega}
%   = \sqrt{\langle\delta\Omega^2\rangle} 
    \propto \bar{N}^{2/3} \,.
\ee
So, the relative width $\Delta_{\Omega}/\bar{\Omega}$ will approach zero, but at a slower pace than when the central limit theorem applies.
    \item Using the method of steepest decent we find that
    \be
        %\mathcal{P}\left( x \right) \stackrel{x \to -\infty}{\approx} \frac{ \sqrt{-x}}{2 \pi }e^{\frac{x^3}{12 \pi }}
        \mathcal{P}\left( x \right) \stackrel{\Omega \ll \bar \Omega}{\approx} \frac{ \sqrt{\bar \Omega-\Omega}}{2 \pi C \bar N}e^{-\frac{(\bar \Omega-\Omega)^3}{12 \pi (C \bar N)^2}}
    \ee
    implying values smaller than the mean, i.e. $\Omega < \bar \Omega$,  are much less likely than for Gaussian distributions -- they are suppressed by an exponent of a cube.
\end{enumerate}

\begin{figure}
\centering
\includegraphics[width=\columnwidth]{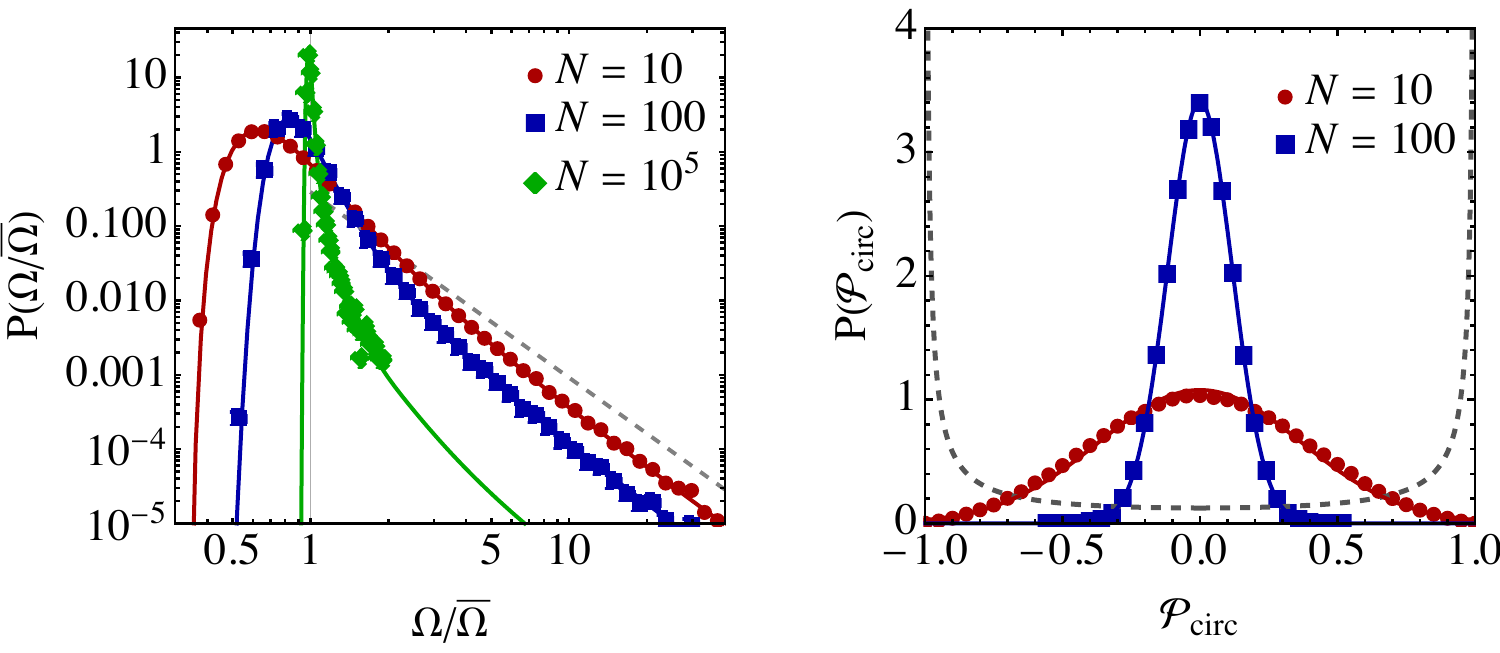}
\caption{{\it Left panel:} The distributions of the total signal strengths from 1, 10, 100, and $10^5$ sources whose independent signal strengths have identical $\Omega^{-\frac52}$ tails. {\it Right panel:} The distribution of $\mathcal{P}_{\rm circ}$ for 1, 10 and 100 identical sources. The dashed line shows the distribution from a single source, the solid lines show the analytic estimates, and the distributions obtained from an explicit Monte Carlo analysis are shown by the points.}
\label{fig:naive}
\end{figure}

\paragraph{An explicit example. --} As an illustration, we consider a population of sources for which individual signals follow the simple long-tailed distribution
\be\label{eq:P1_toy}
    P^{(1)}(\Omega) 
    \propto \Omega^{-\frac{5}{2}} \theta(\Omega - \Omega_{\min})\, ,
\ee
that mimics the statistics of $\Omega_{\rm GW}$ from a realistic BH population found in section \ref{sec:distribution}. In Fig.~\ref{fig:naive}, we show the resulting distributions of $\Omega/\bar\Omega$ if the signal consists of 10, 100, and $10^5$ sources. The mean is given by $\bar{\Omega} \equiv 3N\Omega_{\rm min}$. The distribution is computed by generating random realizations of the source populations (shown by points) and by using \eqref{eq:P_analytic} (shown by solid lines). The two approaches are in excellent agreement. As expected, one can observe that the total signal will inherit the long $\Omega^{-5/2}$ tail from the distribution of individual sources (which is shown by the dashed line). At large $\Omega$, the tail approaches Eq.~\eqref{eq:52tail} in all cases, while the asymptotic \eqref{eq:P1_asympt2} works well for $N = 10^5$, but predicts a slightly too wide peak for $N \leq 10^5$. As predicted, the peak gets narrower and moves closer to the expectation value, when $N$ is increased. We note, however, that although the shape of the simplified distribution in Fig.~\ref{fig:naive} resembles the distribution in Fig.~\ref{fig:MCGW} obtained from a model of the heavy BH binary population, the shape around the peak in the latter depends also on the $z$ dependence and the binary mass distribution.

\subsection{Distribution of $\mathcal{P}_{\rm circ}$} 

Consider now the distribution of the circular polarization
\be
    \mathcal{P}_{\rm circ} \equiv \frac{\sum^{N}_{i=1} V_i}{\sum^{N}_{i=1} I_i} \, ,
\ee
which depends on the distribution and number $N$ of sources. We denote the contributions from individual sources by $V_i$, $I_i$ given by Eqs.~(\ref{eq:hL,R}-\ref{eq:I,V}). As above, we will work in a fixed frequency bin and thus suppress the frequency dependence in the arguments of the functions. 

The polarization will vanish on average, i.e., $\langle\mathcal{P}_{\rm circ}\rangle = 0$, but polarization measurements can provide information via the size of the deviation from that expectation. For a single dominant source, the contribution from the amplitude drops out and we find
\be
    \mathcal{P}_{\rm circ} 
    = \frac{4 c_{\theta}(1+c_{\theta}^2)}{1 + 6c_{\theta}^2 + c_{\theta}^4} \, ,
\ee
where $c_{\theta} \equiv \cos(\theta)$. The corresponding probability distribution is shown as the dashed curve in Fig.~\ref{fig:naive} and can be seen to be dominated by the region $|\mathcal{P}_{\rm circ}| \approx 1$. 

If the signal is dominated by several sources, the analytic treatment of $\mathcal{P}_{\rm circ}$ is complicated due to correlations between $V$ and $I$. The mean and variance of $I$ and $V$ are given by
\bea
    \langle I \rangle &= \frac{4}{5}\langle |h|^2 \rangle \,, 
    \qquad \qquad \
    \langle V \rangle = 0 \,,
    \\
    \langle \delta I^2 \rangle &= \frac{284}{315}\langle (\delta|h|^2)^2 \rangle \,, 
    \ \
    \langle \delta V^2 \rangle = \frac{92}{105} \langle (\delta|h|^2)^2 \rangle\,,
\eea
where we used the fact that the inclination is independent of other parameters of the binary and the fluctuations in $|h|^2$ are computed as in~\eqref{eq:deltaOmega}. The latter is expected to be large. To estimate the fluctuations in $\mathcal{P}_{\rm circ}$ in the limit when the number of sources is large, we first assume that $I$ can be replaced by its average over inclinations $\langle I \rangle_{\theta} \equiv (4/5) \sum_i |h_i|^2$, but not over signal strengths, so that
\be\label{eq:Pcirc_appr}
    \mathcal{P}_{\rm circ} 
    \approx   V/\langle I \rangle_{\theta} \equiv \sum^{N}_{i=1} p_i f_V(c_{\theta,i}) \,,
\ee
where $p_i \equiv (4/5)|h_i|^2/\langle I \rangle_{\theta} \in [0,1]$ is the fractional contribution to the signal from source $i$ and $f_V(c_{\theta}) = (5/4) c_{\theta}(1+c_{\theta}^2)$ contains the inclination dependence of $V$. In this way, it is possible to contain the effect of the large fluctuations in $\langle I \rangle_{\theta}$ in the random fractions $p_i$, which vary only in the range $[0,1]$. We find that
\be\label{eq:deltaPcirc}
    \langle \delta \mathcal{P}_{\rm circ}^2 \rangle
    = \frac{115}{84} N \langle |p_i|^2 \rangle \,.
\ee
%~~\\
We note that $N \langle |p_i|^2 \rangle$ takes values in the range $[1/N,1]$ and could be interpreted as a measure of the effective number of dominant sources $1/N_{\rm dom}$. For instance, in the limiting case where the signal is sourced by $N$ binaries that contribute equally, i.e., $p_i = 1/N$, we would obtain 
\be\label{eq:eq:sigma_circ_2}
    \sigma_{\mathcal{P}_{\rm circ}} = 1.17/\sqrt{N} \, ,
\ee
as was also stated in Eq.~\eqref{eq:sigma_circ}. In the right panel of Fig.~\ref{fig:naive}, approximate Gaussian distributions with a width given by \eqref{eq:eq:sigma_circ_2} (shown by solid lines) are compared to distributions obtained from explicitly generated random populations of $\mathcal{P}_{\rm circ}$ arising from exactly 10 and 100 sources of equal strength (shown by points). We see an excellent agreement between the naive Gaussian approximation and the Monte Carlo estimate already for $N = 10$.

Finally, we remark that the approximation \eqref{eq:Pcirc_appr} above can be improved by expanding $\mathcal{P}_{\rm circ}$ in $\delta I \equiv I - \langle I \rangle_{\theta}$, that is, $\mathcal{P}_{\rm circ} =  V/\langle I \rangle_{\theta}(1 - \delta I/\langle I \rangle_{\theta} + \ldots)$. This allows accounting for correlations between $V$ and $\delta I$, e.g., when computing the variance \eqref{eq:deltaPcirc},
\be
    \langle \delta \mathcal{P}_{\rm circ}^2 \rangle
    = \frac{115}{84} N \langle |p_i|^2 \rangle - \frac{2855}{1386} N \langle |p_i|^3 \rangle + \ldots
\ee
The corrections arising from higher powers of $\delta I/\langle I \rangle_{\theta}$ are suppressed by increasing powers of $1/N$.

\vspace{1cm}

\bibliography{refs}

\end{document}